\documentclass[conference]{IEEEtran}
\pdfoutput=1
\IEEEoverridecommandlockouts
\usepackage{amsmath,amssymb,amsfonts}

\usepackage{algorithm}
\usepackage{algorithmicx}
\usepackage{algpseudocode}
\usepackage{graphicx}
\usepackage{textcomp}
\usepackage{xcolor}
\usepackage{epstopdf}
\usepackage[numbers]{natbib}
\usepackage{subfig}
\usepackage{multicol}
\usepackage{lipsum}
\usepackage{enumerate}

\newtheorem{lemma}{Lemma}
\newtheorem{theorem}{Theorem}

\def\BibTeX{{\rm B\kern-.05em{\sc i\kern-.025em b}\kern-.08em
    T\kern-.1667em\lower.7ex\hbox{E}\kern-.125emX}}
\begin{document}

\title{Simulating Authenticated Broadcast in Networks of Bounded Degree
\thanks{This work has been accepted by ICPADS2021.\\
\\
© 2022 IEEE.  Personal use of this material is permitted.  Permission from IEEE must be obtained for all other uses, in any current or future media, including reprinting/republishing this material for advertising or promotional purposes, creating new collective works, for resale or redistribution to servers or lists, or reuse of any copyrighted component of this work in other works.}
}

\author{\IEEEauthorblockN{1\textsuperscript{st} Shaolin Yu}
\IEEEauthorblockA{\textit{Tsinghua University}\\
Beijing, China \\
ysl8088@163.com}
\and
\IEEEauthorblockN{2\textsuperscript{nd} Jihong Zhu}
\IEEEauthorblockA{\textit{Tsinghua University}\\
Beijing, China \\
jhzhu@tsinghua.edu.cn}
\and
\IEEEauthorblockN{3\textsuperscript{rd} Jiali Yang}
\IEEEauthorblockA{\textit{Tsinghua University}\\
Beijing, China \\
yangjiali-0411@163.com}
\and
\IEEEauthorblockN{4\textsuperscript{th} Yue Ma}
\IEEEauthorblockA{\textit{Tsinghua University}\\
Beijing, China \\
1025583365@qq.com}
}

\maketitle

\begin{abstract}
The authenticated broadcast is simulated in the bounded-degree networks to provide efficient broadcast primitives for building efficient higher-layer Byzantine protocols.
A general abstraction of the relay-based broadcast system is introduced, in which the properties of the relay-based broadcast primitives are generalized.
With this, fault-tolerant propagation is proposed as a building block of the broadcast primitives.
Meanwhile, complementary systems are proposed in complementing fault-tolerant propagation and localized communication.
Analysis shows that efficient fault-tolerant propagation can be built with sufficient initiation areas.
Meanwhile, by integrating fault-tolerant propagation and localized communication, efficient broadcast primitives can be built in bounded-degree networks.
\end{abstract}


\section{Introduction}
Authenticated broadcast is a fundamental building block in constructing easy-understood authenticated Byzantine protocols.
By simulating authenticated broadcast in peer-to-peer networks \citep{SrikanthSimulating}, various kinds of authenticated Byzantine protocols \cite{RN2119,RN4137,Merritt1984Elections,RN4346} can be easily extended to their unauthenticated counterparts in peer-to-peer networks with the broadcast primitive \cite{SrikanthSimulating}.
However, as this broadcast primitive is originally built upon fully connected peer-to-peer networks, its application is limited by the lower bound of the required network connectivity \citep{Dolev1982StrikeAgain}.
With the increasing scale of real-world networks and the restricted independent communication channel resources, the allowed number of faulty nodes often overweighs the allowed node degrees in the peer-to-peer networks.
In this situation, the high network connectivity required in the broadcast primitive \citep{SrikanthSimulating} gravely restricts its application in large networks.

In providing building blocks for Byzantine protocols in large networks, secure communication is viewed as a possible alternative.
By simulating fully connected peer-to-peer communication with secure communication in bounded-degree networks, correct communication can be established between a sufficient number of correct nodes in several synchronous communication rounds.
However, in building higher-layer Byzantine protocols such as Byzantine agreement (BA) with secure communication as the core primitive, the overall message complexity, computational complexity, and required communication rounds are still high.

In this paper, we explore how to simulate authenticated broadcasts in bounded-degree networks effectively.
Firstly, we provide a simple system abstraction of a rich family of relay-based broadcast systems (\emph{broadcast systems} for short), which also includes the original one provided in \cite{SrikanthSimulating} upon fully-connected networks.
With this abstraction, we identify the general properties of the broadcast systems.
Then, to extend the original broadcast system, we investigate the almost everywhere (a.e.) broadcast problem upon bounded-degree networks.
To derive efficient a.e. broadcast solutions, we explore the a.e. propagation problem upon strong enough expanders \citep{RN4359}.
For efficient sublinear-degree broadcast solutions, we investigate the so-called \emph{complementary system} which relatively complements the merits of efficient a.e. propagation and localized communication protocols \citep{Dwork1986,Upfal1992,Chandran2010}.
With the proposed complementary system, more efficient broadcast systems can be built by integrating localized communication protocols and a.e. propagation.
By extending the classical relay-based broadcast system to bounded-degree networks, various Byzantine protocols \citep{SrikanthSimulating,Toueg1987Fast,RN940,RN4084} can be further built upon bounded-degree networks in a simple way.

The rest of this paper is constructed as follows.
The related work and the system model are respectively given in Section~\ref{sec:related} and Section~\ref{sec:model}.
In Section~\ref{sec:Broadcast}, the general broadcast problem is proposed in arbitrarily connected networks.
Then, efficient broadcast solutions are explored in Section~\ref{sec:solution}.
Lastly, we conclude the paper in Section~\ref{sec:con}.

\section{Related work}
\label{sec:related}
In the literature, \cite{SrikanthSimulating} provides the first broadcast primitive that simulates the authenticated broadcast in fully connected reliable peer-to-peer networks.
This primitive facilitates building much easier-understood BA algorithms like \cite{Toueg1987Fast} in comparing with some early explorations like \cite{RN4136}.
In \cite{Daliot2006Agreement}, the broadcast primitive is extended to the bounded-delay model in constructing self-stabilizing BA and other higher-layer real-time protocols \citep{RN940,RN4084}.
However, in real-world communication networks, the reliability of the communication channels and high network connectivity can hardly be both provided.
For example, bus-based networks can simply simulate fully connected networks, but the reliability of the shared communication channels is often low.
Switch-based networks with traffic shaping can provide independence of communication channels, but it is still hard to support high connectivity in real-world applications.
As a result, most large networks are also networks with some bounded node degrees.
In these networks, no broadcast primitive can be applied yet.

As an alternative, secure communication is proposed as a core primitive in building higher-layer Byzantine protocols in bounded-degree networks.
In \cite{Dwork1986}, \emph{almost everywhere} Byzantine protocols (or saying \emph{incomplete} Byzantine protocols \citep{BPG1989} in a broader meaning) are first intuitively provided upon some constant-degree networks.
However, the original a.e. Byzantine solution \citep{Dwork1986} tolerates only $O(n/\log n)$ faults even when $n$ is much small.
In \cite{Upfal1992}, the a.e. Byzantine solutions can tolerate a linear number of faults with a linear number of poor nodes upon some constant-degree networks.
However, some fault-tolerant operation with very high computational complexity is demanded.
In \cite{Chandran2010}, the computational complexity and the number of the poor nodes are both asymptotically reduced by taking a multi-layer transmission scheme and allowing the node-degree to be polylogarithmic.
However, the communication network constructed corresponding to the specially designed multi-layer transmission scheme is rather complex and lacks simplicity.
Also, in running the transmission scheme, each node needs not only to transmit the passing messages but run some sub-layer fault-tolerant protocols \citep{Garay2003} for the passing messages, which still generates considerable computation, time, and message complexities.
In \cite{Jayanti2020}, it is shown that there exist more efficient transmission schemes and communication networks with allowing polylogarithmic-degrees, but the construction of such networks is not explicit nor deterministic yet.
Also, in considering the overall efficiency, all these a.e. Byzantine solutions aim only at secure communication between the so-called \emph{privileged} nodes.
In constructing upper-layer Byzantine protocols like BA, the time needed to execute the low-layer communication protocol is often a factor of the overall execution time.
In this sense, the overall complexity of the secure-communication-based deterministic BA is at least polynomial.
In breaking these barriers, only probabilistic solutions are further investigated \citep{BENOR1996329boundeddegree,King2011Breaking}.

\section{The system model}
\label{sec:model}
The synchronous system $\mathcal{S}$ consists of $n$ nodes, denoted as $V$ (let $V$ being represented as $\{1,2,\dots,n\}$ for convenience), in which up to $f=\alpha n$ nodes can fail arbitrarily (we assume $0\leqslant\alpha<1$ and ignore all the trivial rounding problems).
All nodes other than the faulty ones are correct.
The bidirectional connections between the $n$ nodes are represented as the edge-set $E$ of the undirected graph $G=(V,E)$.

During each basic synchronous round (\emph{round} for short), each correct node $i\in V$ can send one or more messages to all its neighbor nodes (denoted as $N_i$ and we assume $i\in N_i$ for convenience), receive all the valid messages sent from $N_i$ during the same round and complete all needed process according to the provided algorithms before the beginning of the next round.
In any round, a faulty node $i'$ can send arbitrarily inconsistent valid messages, invalid messages, or nothing to any subset of $N_{i'}$.
For simplicity and without loss of generality, here we always assume that in each round, each correct node $i$ would distribute a valid message $m$ contains a value $v\in \mathbb V$ that can be correctly extracted in every correct node in $N_i$ during the same round.
In the basic system settings, we set $\mathbb V =\{0,1\}$.
In this case, when a node sends no message to a correct node in a round, we assume the corresponding value would be extracted as $0$.
And whenever a correct node $i$ extracts a value $v\neq 0$ from any node (might be faulty), $i$ would set $v$ as $1$.
Thus, when a correct node sends a message $m$, we can always assume $m$ contains the value $1$.
We can see this would simplify the basic discussions.

For all the cases, we assume the faulty nodes being under the complete control of a malicious \emph{strong} adversary who knows \emph{everything} of the system.
Namely, this \emph{strong} adversary knows the network topology $G$, the algorithms provided for the correct nodes in $\mathcal{S}$, the initial state of $\mathcal{S}$ and all events generated during every execution of $\mathcal{S}$ even before these events being generated.
With this, the adversary can arbitrarily select any subset $T\subset V$ from $V$ with $|T|\leqslant f$ at the first round and send arbitrary messages from $T$ during each round in every execution of $\mathcal{S}$.

We say $\mathcal{S}$ is a broadcast system upon $G$ if and only if (iff) $\mathcal{S}$ can simulate the authenticated broadcast \citep{SrikanthSimulating} upon $G$.
In measuring the Byzantine resilience, $\mathcal{S}$ is an $\alpha$-resilient broadcast system iff the desired authenticated broadcast can be simulated in all correct nodes in the presence of $\alpha n$ Byzantine nodes.
For bounded-degree networks, $\mathcal{S}$ is an $(\alpha,\mu)$-resilient incomplete broadcast system with $1<\mu<\alpha^{-1}$ iff desired authenticated broadcast can be simulated in at least $(1-\mu\alpha) n$ correct nodes.
For convenience, an $\alpha$-resilient broadcast system is also the $(\alpha,\mu)$-resilient broadcast system with $\mu=1$.
It should be noted that although we allow $\mu>1$ in the incomplete systems, all the systems discussed in this paper are deterministic, i.e., we consider only the solutions for the worst cases.

With these, the problem is to establish the broadcast system upon the bounded-degree network $G$.

\section{The broadcast problem}
\label{sec:Broadcast}
In this section, we first extend the broadcast problem under a general system structure, with which the broadcast systems upon arbitrarily connected networks can be further explored.

\subsection{A general system structure}
\begin{figure}[htbp]
\centering
\subfloat[Abstraction\label{fig:local_system1}]{\centering\includegraphics[width=1.0in]{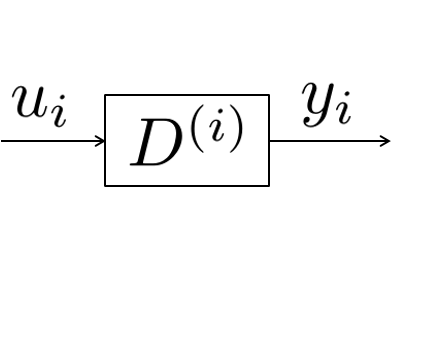}}
\subfloat[The Structure\label{fig:local_system2}]{\centering\includegraphics[width=2.3in]{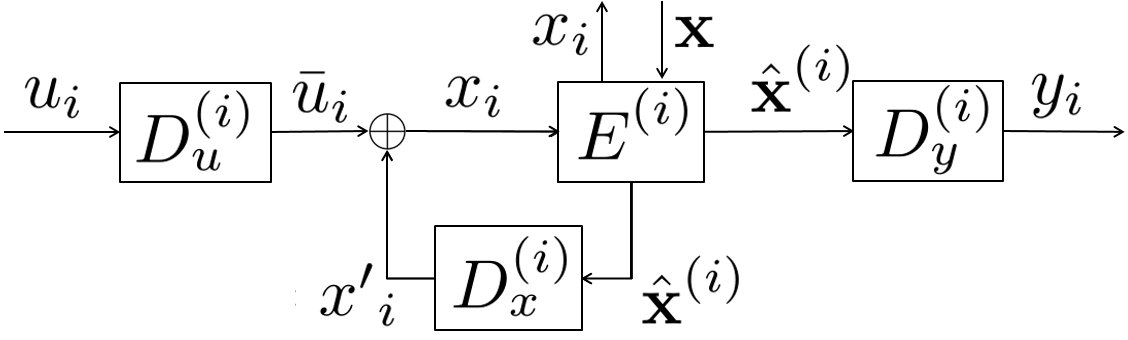}}
\caption{A Local System.}
\end{figure}

In the broadcast system $\mathcal{S}$, each node $i\in V$ can be viewed as a local system $D^{(i)}$ running on the discrete-time $k\in \mathbb Z$.
In the context where only one execution of the broadcast system is considered, the discrete-time $k$ can be directly viewed as the round numbers in $\mathbb N=\{0\}\cup \mathbb Z^+$, i.e., we can interchange the word \emph{time} and \emph{round} in this context.
With the discrete time $k$, a \emph{signal} is defined as a function $s:\mathbb Z \to \mathbb V$ which gives a unique value $s(k)\in \mathbb V$ for each $k\in \mathbb Z$.
With this, the input signal of $D^{(i)}$ corresponds to the extracted message values $u_i(k)$ that come from the broadcaster (also referred to as the \emph{General}).
The output signal of $D^{(i)}$ corresponds to the yielded decision values $y_i(k)$, as is shown in Fig.~\ref{fig:local_system1}.

We say $D^{(i)}$ is correct iff $i\in U$.
To support the desired properties of the broadcast system, a correct $D^{(i)}$ also generates a signal $x_i$ during each execution.
By definition, this signal is intended to indicate other local systems about the current local state $x_i(k)$ of node $i$ in round $k$.
In the provided broadcast system upon $K_n$ \citep{SrikanthSimulating}, the signal $x_i$ and $y_i$ are all monotonically increasing.
For simplicity, in the general broadcast systems, we still assume the signal $x_i$ and $y_i$ of the correct $D^{(i)}$ are monotonically increasing.

To process the signals, a correct $D^{(i)}$ is composed of several basic blocks, as is shown in Fig.~\ref{fig:local_system2}.
Generally, the block $D_u^{(i)}$ transfers the raw \emph{General} input $u_i(k)$ into a valid value $\bar{u}_i(k)\in \mathbb V$ in node $i$ at round $k$.
Then $\bar{u}_i(k)$ is added to the temporally computed local state ${x'}_i(k)=D_x^{(i)}(\hat{\vec{x}}^{(i)}(k-1))$ in current round $k$ to update the current local state of node $i$ as $x_i(k)={x'}_i(k)+\bar{u}_i(k)$.
The vector $\hat{\vec{x}}^{(i)}(k-1)$ is the previous round estimation of system state in node $i$.
The current local state $x_i(k)$ in each node $i$ is exchanged with that of the neighbor nodes $N_i$ in block $E^{(i)}$ to collect an estimation of current system state in each node $i$ as $\hat{\vec{x}}^{(i)}(k)$.
As the network $G$ can be arbitrarily connected, the block $E^{(i)}$ can only output the states of the neighbours of node $i$ in $G$.
Then, the block $D_y^{(i)}$ transfers this $\hat{\vec{x}}^{(i)}(k)$ into the decision value $y_i(k)=D_y^{(i)}(\hat{\vec{x}}^{(i)}(k))$.
In the correct $D^{(i)}$, $D_u^{(i)}$, $D_x^{(i)}$ and $D_y^{(i)}$ are all stateless and can only output values in $\mathbb V$.

\begin{figure}[htbp]
\centerline{\includegraphics[width=2.6in]{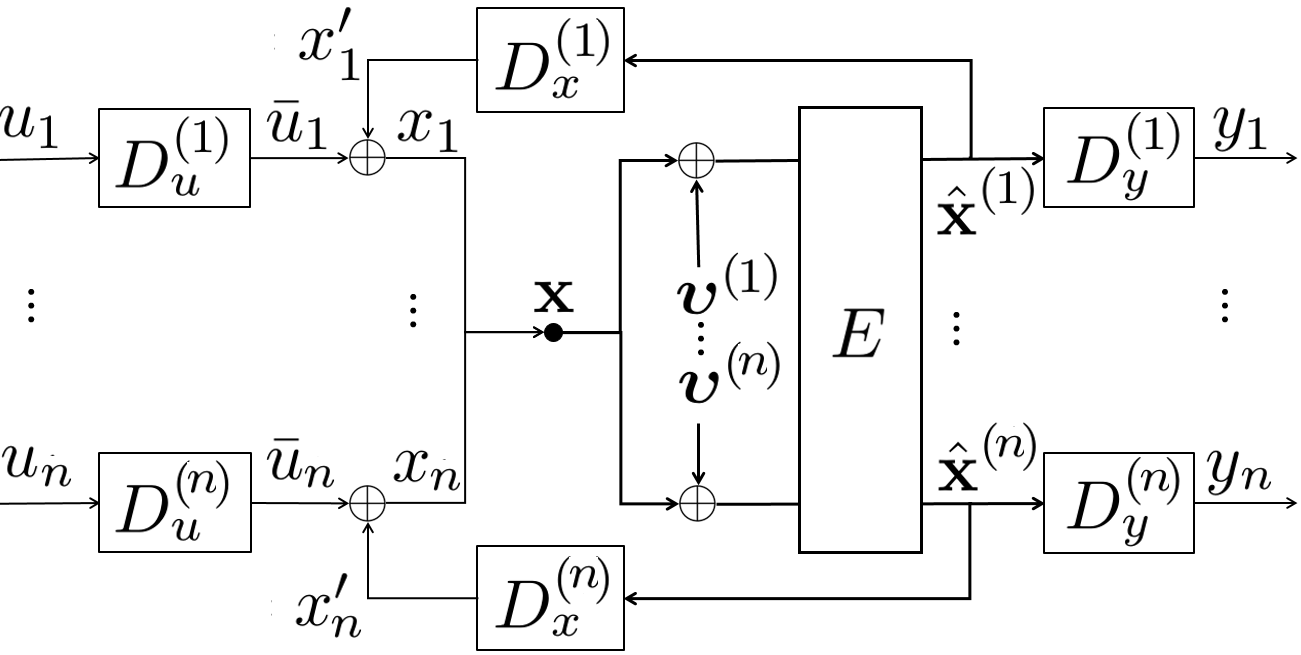}}
\caption{The Generalized Broadcast System upon $G$.}
\label{fig:broadcast_system}
\end{figure}

Meanwhile, a faulty (i.e., not correct) local system $D^{(i')}$ can generate not only arbitrary local state $x_{i'}(k)$ but also make $x_{i'}(k)$ being inconsistently measured as ${x'}^{(i)}_{i'}(k)$ in the correct nodes $i\in N_{i'}$.
In this sense, there might be no actually unique system state in any round in considering the \emph{multi-faced} faulty local systems.
But equivalently, we can always assume that a unique system state ${\vec{x}}(k)=\langle x_1(k),\dots,x_n(k)\rangle$ is first generated by all $n$ correct local systems and is then interfered by some noises ${\vec{\upsilon}}^{(i)}(k)=\langle \upsilon^{(i)}_1(k),\dots,\upsilon^{(i)}_n(k)\rangle$ with $\upsilon^{(i)}_j(k)\in \mathbb{F}=\{0,1,-1\}$ before it entering $E^{(i)}$ at round $k$.
Similarly, the decision vector ${\vec{y}}(k)=\langle y_1(k),\dots,y_n(k)\rangle$ can also be assumed being yielded by $n$ correct local systems.
In other words, a faulty node $i'$ still has the chance to behave as a correct node whenever it likes.
For simplicity, we assume $\bar{\vec{u}}\equiv {\vec{u}}$ and the signals ${\vec{u}},{\vec{y}}$ are extended to negative time (when $k<0$) with value $\vec0$.
In this manner, the general broadcast system upon $G$ is represented as the system $\mathcal{D}$ whose structure is shown in Fig.~\ref{fig:broadcast_system}.

\subsection{Basic equations}
\label{subsec:basic-transition-functions}
With this, we can represent $D^{(i)}$ for every $i\in V$ as
\begin{eqnarray}
\label{eq:functions_transfer}&&{x}_{i}(k)=D_x(\hat{\vec{x}}^{(i)}(k-1))+D_u({u}_{i}(k)) \\
\label{eq:functions_estimate}&&\hat{\vec{x}}^{(i)}(k)=E^{(i)}(\vec{x}(k)+\vec{\upsilon}^{(i)}(k)) \\
\label{eq:functions_yield}&&{y}_{i}(k)=D_y(\hat{\vec{x}}^{(i)}(k))
\end{eqnarray}
where $\hat{\vec{x}}^{(i)}(k)$ is the estimated system state in node $i$ at round $k$ and the functions $D_x$, $D_y$ and $D_u$ are uniformly defined in all nodes (as only uniform solutions are considered).
Here, the operator $+$ is a function from $\mathbb{F} \times \mathbb{F}$ to $\mathbb{V}=\{0,1\}$, with $a+b=0$ iff $a=-b$.

In every execution of the broadcast system, as there can be \emph{multi-faced} faulty local systems, the noise vectors $\vec{\upsilon}^{(i)}(k)$ in (\ref{eq:functions_estimate_general}) for different $i_1,i_2\in V$ can be different with the same $k$.
In such an execution, the noises $\langle \vec{\upsilon}^{(1)}(k),\dots,\vec{\upsilon}^{(n)}(k) \rangle$ in round $k$ can be represented as an $n\times n$ noise matrix $F(k)$.
The $i$th column vector of $F(k)$ represents a state noise vector ${\vec{\upsilon}}^{(i)}(k)$ measured in node $i$.
And if node $j$ is correct, the $j$th row vector in $F(k)$ equals to $\vec{0}$.
Otherwise, if node $j$ is faulty, the $j$th row vector in $F(k)$ can have arbitrary values in $\mathbb F$.
These arbitrarily valued rows in $F(k)$ are called Byzantine rows.
As there can be up to $f$ faulty nodes in round $k$, up to $f$ Byzantine rows can be scattered on $F(k)$ in all possible combinations.
Here, a noise matrix $F$ with up to $f$ Byzantine rows is referred to as an $f$-Byzantine matrix $F^{[f]}$.
The set of all $f$-Byzantine matrices are denoted as $\Upsilon^{[f]}$.
By this definition, we have $F^{[f_0]}\in \Upsilon^{[f]}$ when $f_0\leqslant f$.

With this, the broadcast system $\mathcal{D}$ can be represented as
\begin{eqnarray}
\label{eq:functions_transfer_general}&&\vec{x}(k)=D_x(\hat{\vec{x}}(k-1))+D_u(\vec{u}(k)) \\
\label{eq:functions_estimate_general}&&\hat{\vec{x}}(k)=E\odot ({\vec 1}^T \otimes \vec{x}(k)+F(k))\\
\label{eq:functions_yield_general}&&\vec{y}(k)=D_y(\hat{\vec{x}}(k))
\end{eqnarray}
where $\hat{\vec{x}}(k)=[ \hat{\vec{x}}^{(1)}(k),\dots,\hat{\vec{x}}^{(n)}(k)]$, $F(k)\in \Upsilon^{[f]}$, $E=I_n+A$ (here we slightly abuse the edge-set $E$ as also a matrix for convenience when it is not confusing), $A$ is the adjacency matrix of $G$, $I_r$ is the $r\times r$ identity matrix with $r\geqslant 1$, $\otimes$ is the Kronecker product, and the mask operator $\odot $ always computes a matrix with elements $m_{i,j}=p_{i,j}q_{i,j}$ for the same-sized matrices, where $x_{i,j}$ denotes the element in the $i$th row and $j$th column of a matrix.
As a special case, the broadcast system upon $K_n$ can be viewed as $\mathcal{D}$ with $E=J_{n,n}$, where $J_{r,s}$ is the $r\times s$ all-ones matrix.
As the system is distributed, $D_u$, $D_x$, and $D_y$ can use only information in each column of the function-input in computing the corresponding value of each element in the function-output vectors.

Denoting the set of all possible executions of $\mathcal{D}$ as $\Lambda_{D}$, if all noise matrices in an execution $\chi\in \Lambda_{D}$ are in $\Upsilon^{[f]}$, $\chi$ is referred to as an $f$-Byzantine execution, denoted as $\chi\in \Lambda_{D}^{[f]}$.
For different $\chi_1,\chi_2\in \Lambda_{D}^{[f]}$, the output signals in $\mathcal{D}$ can be different as the input signals and the noises can all be different.

\subsection{Required properties}
In the broadcast system upon complete graph $K_n$, if the correct \emph{General} initiates a broadcast at time $k_0$, it would always generate a valid input signal in every $D^{(i)}$ as $u_i(k)\equiv \delta[k-k_0]$ where $\delta$ is the discrete Dirac function (i.e., with $\delta[0]=1$ and $\delta[k]=0$ for all $k\neq 0$).
In this case, with the \emph{correctness} property, every correct local system $D^{(i)}$ should yield the decision signal $y_i(k)\equiv H[k-k_0]$, where $H$ is the corresponding discrete Heaviside step function (i.e., $H[k]\equiv\sum_{ k'=0}^{k}\delta[k']$).
For convenience we also denote $s_{+k_0}$ as the signal $s_{+k_0}(k)\equiv s(k-k_0)$.
With the \emph{unforgeability} property, if the correct \emph{General} does not initiate a broadcast before $k_0$, every correct $D^{(i)}$ cannot yield a decision signal $y_i(k)\equiv H[k-k']$ with any $k'<k_0$.
Thus, for a correct \emph{General}, it requires
\begin{equation}
\label{eq:Heaviside}
(\exists u:\forall i\in U :u_i= u)\to (\forall i\in U :y_i=\sum u)
\end{equation}
where $\sum s$ is the integral of the signal $s$.
This is referred to as the $1$-\emph{Heaviside integral} ($1$-\emph{Heaviside} for short) property of $\mathcal{D}$.
Otherwise, if the \emph{General} is faulty, with the \emph{relay} property, in every $\chi\in \Lambda_{D}^{[f]}$ it requires
\begin{equation}
\label{eq:Dirac}
\exists k'\geqslant 0: \forall i,j\in U:\left | y_i-y_j \right |\leqslant \delta_{+k'}
\end{equation}
holds, where $|s|$ is naturally defined as the absolute-value signal of the signal $s$ (i.e., $|s|(k)\equiv |s(k)|$ holds, and all other operators on the signals are naturally defined similarly).
In other words, the $y_i$ signals yielded in all correct $D^{(i)}$ are allowed up to one $\delta$ apart (viewed as the $\delta$-distance of the signals).
This is referred to as the \emph{1-Dirac differential} (\emph{1-Dirac} for short) property of $\mathcal{D}$ (Note that the Dirac differential property here should not be confused with that of the $\delta$-differential consensus in \cite{Garay2003} where $\delta$ describes a property of the initial values of the consensus).

In the broadcast system upon the arbitrarily connected $G$, there are differences.
Firstly, if the correct \emph{General} initiates a broadcast at $k_0$, not all correct local systems can be input with $\delta_{+k_0}$.
Instead, only the local systems in some $I_0  \subseteq V$ can be initiated by the correct \emph{General}.
In this situation, we always have $\|\vec u(k_0)\|\leqslant|I_0  |$, where ${\lVert \vec{r} \rVert}$ is the $0$-norm of a vector $\vec{r}$ (i.e., the number of nonzero elements in $\vec{r}$).
More specifically, defining $b_{a,A}=1$ if $a\in A$ and $b_{a,A}=0$ otherwise, we have $u_i=b_{i,I_0  }\delta_{+k_0}$ when the \emph{General} is correct and $u_i=c_i b_{i,I_0  }\delta_{+k_0}$ with $c_i\in \mathbb V$ being arbitrarily valued when the \emph{General} is not correct.
Here, $I_0  $ is called an initiation set upon $G$.
The set of all initiation sets upon $G$ is denoted as $\mathbf{I}_G$.

Secondly, in this case, with the \emph{extended correctness and unforgeability} property, for every $\chi\in \Lambda_{D}^{[\alpha n]}$ it requires
\begin{eqnarray}
\label{eq:Heaviside_ex}
\forall I_0  \in \mathbf{I}_G:(\exists u:\forall i\in V :u_i= b_{i,I_0  }u)\to\nonumber \\
(\exists P\subseteq U: |P|\geqslant (1-\mu\alpha)n \land \forall i\in P : \nonumber \\
\exists 0\leqslant k_1<k_H : y_i=\sum u_{+k_1})
\end{eqnarray}
with a bounded $k_H$ and sufficiently small $\mu$.
This is referred to as the $k_H$-\emph{Heaviside} property of an $(\alpha,\mu)$-resilient $\mathcal{D}$.

Thirdly, with the \emph{extended relay} property, for every $\chi\in \Lambda_{D}^{[\alpha n]}$ it requires
\begin{eqnarray}
\label{eq:Dirac_ex}
\exists P\subseteq U, k_1\leqslant\dots\leqslant k_m< k_1+k_\delta: |P|\geqslant (1-\mu\alpha)n \land  \nonumber \\
 \forall i,j\in P:| y_i-y_j |\leqslant \sum_{r=1}^{m}\delta_{+k_r}
\end{eqnarray}
with a bounded $k_\delta$.
And this is referred to as the $k_\delta$-\emph{Dirac} property of an $(\alpha,\mu)$-resilient $\mathcal{D}$.

With this, $\mathcal{D}$ is an $(\alpha,\mu)$-resilient $(k_H,k_\delta)$ broadcast system upon $G$ iff the $(\alpha,\mu)$-resilient $\mathcal{D}$ satisfies the $k_H$-\emph{Heaviside} and $k_\delta$-\emph{Dirac} properties with all initiation sets in $\mathbf{I}_G$.

\section{The broadcast systems upon $G$}
\label{sec:solution}
One significant problem with the complete network is that the node-degrees are linear to $n$ and thus are not bounded.
Practically, with the increase of $n$, it is crucial to maintain the required node-degrees of the network within some affordable scale.
In this section, we investigate the $d$-regular networks with $f=\alpha n$, where $\alpha>0$ should be independent of $n$, and $d$ should be sublinear to $f$.
We can see that these requirements exclude some natural solutions, such as the ones allowing $d=\Omega(f)$.
However, this requirement also naturally comes from the real world.
Firstly, with the increasing numbers of unreliable components in distributed systems, the allowed numbers of faulty components should be increased accordingly.
As real-world common networking products are always with a restricted number of communication channels, $d$ should remain affordable despite increasing system scales.

\subsection{The fault-tolerant propagation upon $G$}
Under the system structure of $\mathcal{D}$, when a correct node $i\in V_0$ for some $V_0\in \mathbf{I}_G$ is input with the Dirac signal $\delta$, the local state $x_i$ should be set as the Heaviside signal according to the monotonic assumption.
This can be viewed as the node $i$ being excited by the input $\delta$.
From this point on, the execution of $\mathcal{D}$ upon $G$ can be intuitively viewed as the propagation of the excitation signals in some excitable media \citep{RN4360} with the topology $G$.
In satisfying the $k_H$-\emph{Heaviside} property required in (\ref{eq:Heaviside_ex}), the excitation signals of an arbitrary initiation set should be propagated to at least $(1-\mu\alpha)$ area of the whole excitable media within at most $k_H$ discrete-time in the presence of an arbitrarily distributed $\alpha$ area being arbitrarily faulty.
Besides, the propagation should be prevented in such a $(1-\mu\alpha)$ area when there is no initial excitation in this area.
And to satisfy the $k_\delta$-\emph{Dirac} property required in (\ref{eq:Dirac_ex}), whenever the output signal $y_i$ of an npc node $i$ is triggered at time $k_0$ (i.e., $y_i\equiv H_{+k_0}$), the output signals of the correct nodes in at least $(1-\mu\alpha)$ area of $G$ should also be triggered before $k_0+k_\delta$, with which all the npc nodes would be triggered.
Here, the natural idea is first to design some desired propagation protocol, denoted as $\mathcal{P}$, upon which the desired $k_H$-\emph{Heaviside} and $k_\delta$-\emph{Dirac} properties can be built after that.

Firstly, to conveniently observe the desired propagation, we can always rearrange the order of the $n$ nodes at any time $k$ to make the state vector $\vec{x}(k)$ of $\mathcal{D}$ being in the form $\langle 1,\dots,1,0,\dots,0\rangle$, i.e., the excited nodes always with smaller index than the unexcited ones.
With this, the $n^2$ elements in $A$ (the adjacency matrix of $G$) can be rearranged accordingly.
As $\vec{x}$ is monotonic, we can make the matrix $E=I_n+A$ being constant with respect to the time $k$ during any execution of $\mathcal{D}$.
So with (\ref{eq:functions_estimate_general}), we have
\begin{eqnarray}
\label{eq:func_rearrange}
\hat{\vec{x}}(k)=E\odot X(k)+E\odot F(k)
\end{eqnarray}
where the $n\times n$ matrix $X(k)$ is in the form $[J_{n,m}~0]^T$ with $m=\|\vec{x}(k)\|$.
So $X(k)$ can be viewed as a mask matrix for $E$ that always propagates $\vec 1$ from the left side to the right side.

To satisfy the $k_\delta$-\emph{Dirac} property, together with (\ref{eq:functions_transfer_general}) and (\ref{eq:functions_yield_general}), when $\|\vec{x}(k_0)\|\geqslant \min\{\|\vec{x}\|\mid D_y(\vec{x}+\vec{\upsilon})=1\}$, we require
\begin{eqnarray}
\label{eq:necessary_propa0}
\|\vec{x}(k+1)\|=\|D_x(E\odot X(k)+E\odot F(k))\|\geqslant~~~~\nonumber\\
 \min\{\|\vec{x}(k)\|+1,(1-\mu\alpha)n\}~~~~
\end{eqnarray}
holds for all $k\geqslant k_0$ in all executions of $\mathcal{D}$.
Further, in uniform solutions where the nodes are all equally weighted in all correct nodes, this condition can be simplified as
\begin{eqnarray}
\label{eq:necessary_propa}
\exists k_0:D_y(E\cdot(X(k_0)+F(k_0)))\neq 0\to \nonumber\\
\forall k\geqslant k_0:\|D_x(E\cdot X(k)+E\cdot F(k))\|\geqslant \nonumber\\
\min\{\|\vec{x}(k)\|+1,(1-\mu\alpha)n\}
\end{eqnarray}
where $\cdot$ is the common matrix multiplication operator and $D_x,D_y$ are all Heaviside functions.
We can see that this is just the natural extension of the relay strategy (proposed in \cite{SrikanthSimulating}) for satisfying the $k_\delta$-\emph{Dirac} property upon $G$.

To satisfy the $k_H$-\emph{Heaviside} property, we still require that when no correct node is initially excited in some node-set $P\subseteq U$ with $|P|\geqslant (1-\mu\alpha)n$, then no node in $P$ would be excited during the propagation.
It should be noted that even under such simplification, the problem of satisfiability of (\ref{eq:necessary_propa}) and the $k_H$-\emph{Heaviside} property upon the generally connected network $G$ is still nontrivial.

\subsection{A sufficient condition for a.e. $\epsilon$-incomplete propagation}
For large-scale systems, \cite{Upfal1992} shows that by explicitly constructing a $d$-regular Ramanujan network $G$ (i.e., with $\lambda=\max\{|\lambda_i|\}\leqslant 2\sqrt{d-1}$ where $\{\lambda_i\}$ are the eigenvalues of the adjacency matrix of $G$ in the $(-d,d)$ interval \citep{RN4359}) \citep{ALON198815} with a sufficiently large $d$, $O(n)$ Byzantine faults can be tolerated in reaching a.e. $\epsilon$-BA among the $n$ connected $d$-degree nodes.
Concretely, following \cite{RN4359}, for any two primes $p\equiv q\equiv 1 \bmod 4$ with Legendre symbol $(\frac{p}{q})=-1$, we can construct a $(p+1)$-regular bipartite Ramanujan network with $n=q(q^2-1)$ nodes.
And when $(\frac{p}{q})=1$, a \emph{non-bipartite} Ramanujan network with $n=q(q^2-1)/2$ nodes can also be constructed.
Upon this, the basic fault-tolerant strategies provided in \cite{Upfal1992} can be employed in the $d$-regular non-bipartite networks for any $d\geqslant p+1$.
Here for simplicity, we assume $d=p+1$ and first aim for providing an a.e. $\epsilon$-broadcast system upon $d$-regular \emph{non-bipartite} Ramanujan networks (also referred to as Ramanujan networks).

Firstly, it is known that the \emph{non-bipartite} Ramanujan networks have the following basic property \citep{ALON198815}.
\begin{lemma}[\citep{ALON198815}]
\label{lemma_ramanujan}
If $G=(V,E)$ is a connected non-bipartite Ramanujan network, then for every $S\subseteq V$ with $|S|=\theta n$
\begin{eqnarray}
\label{eq:ramanujan}
|e(S) -\theta^2dn/2|\leqslant \sqrt{d-1}\theta(1-\theta)n
\end{eqnarray}
holds, where $e(S)=|E\cap (S\times S)|$ is the number of the internal edges of the subgraph of $G$ induced by $S$.
\end{lemma}
\begin{IEEEproof}
As $G$ is a connected non-bipartite Ramanujan network \citep{RN4359}, $d$ is a simple eigenvalue of $A$ (the adjacency matrix of $G$) and the absolute-values of all the other $n-1$ eigenvalues of $A$ are no more than $2\sqrt{d-1}$.
So the conclusion holds with Lemma~2.3 of \cite{ALON198815}.
\end{IEEEproof}

Now to provide the desired system, in ease of the propagation of the excitation signals in any $V_0\in \mathbf{I}_G$, $|V_0|$ should be as large as possible, and the propagation condition (such as the threshold function $D_x=H_{+m}$) should be as loose as possible.
However, in $d$-regular networks, it is impractical to require $|V_0|$ being larger than $d+1$ in the absence of the underlying communication protocol.
Meanwhile, in considering Byzantine faults, it is also nonsense to require $m\leqslant 1$.
In this situation, let $m=\beta d+1$ in the threshold function $D_x=H_{+m}$, where $\beta\in (0,1)$ is called the propagation coefficient.
Namely, a correct node $i\in U$ would be excited at time $k$ iff $i$ receives at least $\beta d$ excitation signals from the $1$-neighbors of $i$ (defined as $N_i\setminus i$) at some $k_0\leqslant k$.
Similarly, let $D_y=H_{+(\beta_2 d+1)}$ where $\beta_2\in (0,1)$ is called the triggering coefficient.
Now, we show that by taking a sufficiently large $d$, the propagation can reach almost everywhere of the excitable media when the initial excited area is relatively small.
Meanwhile, when almost everywhere of the excitable media is not excited, it remains to be unexcited.

Firstly, the $\mathbf{P}$ function introduced in \cite{Upfal1992} can be generalized to construct the smallest node-set $Z$ with satisfying $T\subseteq Z$ and $\{i\in V\mid |N_i\cap Z|\geqslant \beta_0 d\}\subseteq Z$ for every $T$, where $\beta_0\in (0,1)$ is called the immunity coefficient.
Here we assume $G=(V,E)$ is an $n$-vertex $d$-regular non-bipartite Ramanujan network and denote such constructed $Z$ as ${Z}(T,\beta_0)$ and the set of the npc nodes as ${P}(T,\beta_0)=V\setminus({Z}(T,\beta_0)\cup T)$ just following \cite{Upfal1992}.
Then Lemma~1 of \cite{Upfal1992} can be generalized with $\beta_0$ as follows.

\begin{lemma}
\label{lemma_sufficient_large0}
For any $\alpha,\beta_0\in (0,1)$, if
\begin{eqnarray}
\label{eq:necessary_large}
\beta_0-\sqrt{2\alpha\beta_0}\geqslant\sqrt{d-1}/d
\end{eqnarray}
then there exists $\mu< \sqrt{2\beta_0/\alpha}$, such that $\forall T\subset V: |T|\leqslant\alpha n \to |P(T,\beta_0)|> n-\mu |T|$.
\end{lemma}
\begin{IEEEproof}
Let $|{Z}(T,\beta_0)\cup T|=\mu_0 |T|$.
Then for every $\mu\in(1,\mu_0)$, as the subgraph of $G$ induced by any $S\subseteq{Z}(T,\beta_0)\cup T$ with $|S|=\mu |T|$ has at least $(\mu-1)|T|\beta_0 d$ internal edges, with Lemma~\ref{lemma_ramanujan}, $|\beta_0(\mu-1)/\mu-\alpha\mu /2|<\sqrt{d-1}/d$ holds.
Denote $g(x)=\beta_0(x-1)/x-\alpha x /2$ and suppose $\mu_0\geqslant\sqrt{2\beta_0/\alpha}$.
As $g(\sqrt{2\beta_0/\alpha})=\beta_0-\sqrt{2\alpha\beta_0}$, $\beta_0-\sqrt{2\alpha\beta_0}<\sqrt{d-1}/d$ holds for $\mu=\sqrt{2\beta_0/\alpha}$.
A contradiction.
\end{IEEEproof}

It is somewhat weird to see $\mu$ being taken as $\sqrt{2\beta_0/\alpha}$, which apparently says that a smaller $\beta_0$ promises a smaller $\mu$.
This is because that such $\mu$ is taken as the peak of the $g$ function defined in Lemma~\ref{lemma_sufficient_large0}.
Actually, the nontrivial lower-bound of $\beta_0$ is restricted by (\ref{eq:necessary_large}), where $\beta_0$ is required to be sufficiently large to make the edges between the nodes in ${Z}(T,\beta_0)\cup T$ being sufficiently dense.
Also, note that in Lemma~\ref{lemma_sufficient_large0}, an implicit condition is $\sqrt{2\beta_0/\alpha}>1$.
This can also be deduced from the condition (\ref{eq:necessary_large}) required in Lemma~\ref{lemma_sufficient_large0}.
Now, to satisfy (\ref{eq:necessary_large}), as $\lim_{d\to +\infty}\sqrt{d-1}/d=0$, i.e., for every $\epsilon_0>0$ there exists $d>0$ making $\sqrt{d-1}/d\leqslant\epsilon_0$, a sufficiently large $d$ would do iff $\beta_0>\sqrt{2\alpha\beta_0}$, for which $\beta_0>2\alpha$ should be satisfied.

Now we show that the $\epsilon$-incomplete propagation can be accomplished upon this Ramanujan network $G$ if the initiation set is affordable, providing that $\beta$ is sufficiently small.

\begin{lemma}
\label{lemma_propagating}
For any $\alpha,\beta,\beta_0,\beta_2,\theta_0\in (0,1)$, if
\begin{eqnarray}
\label{eq:necessary_small}
\beta+3(\beta_0-\sqrt{2\beta_0\alpha})<\theta_0
\end{eqnarray}
and the inequality (\ref{eq:necessary_large}) hold and $\exists P_1\subseteq {P}(T,\beta_0):|P_1|\geqslant \theta_0 n \land \forall i\in P_1:x_i(k_0)=1$, then $\forall i\in{P}(T,\beta_0): x_i(k)=1$ holds for all $k\geqslant k_0+|{P}(T,\beta_0)|$ in $\mathcal{D}$ upon $G$.
\end{lemma}
\begin{IEEEproof}
Inspired by \cite{Upfal1992}, now suppose that there exists $S\subseteq {P}(T,\beta_0)$ with $|S|=\theta n$ and $\forall i\in S:x_i=0 \land |N_i\cap ({P}(T,\beta_0)\setminus S)|<\beta d$.
Then there are less than $\beta d|S|$ edges between $S$ and ${P}(T,\beta_0)\setminus S$.
As each node $i\in S$ has more than $(1-\beta_0)d$ $1$-neighbors in ${P}(T,\beta_0)$, the subgraph of $G$ induced by $S$ has more than $((1-\beta_0)d|S|-\beta d|S|)/2=(1-\beta-\beta_0)d|S|/2$ internal edges.
With Lemma~\ref{lemma_ramanujan}, $|(1-\beta-\beta_0)d\theta n/2-d\theta^2n/2|\leqslant \sqrt{d-1}\theta(1-\theta)n<\sqrt{d-1}\theta n$ holds.
So $1-\beta-\beta_0-\theta<2\sqrt{d-1}/d$ should be satisfied.
Now with the existence of $P_1$, we have $|S|\leqslant |{P}(T,\beta_0)|-|P_1|\leqslant (1-\mu\alpha)n-\theta_0 n)$ for $\mu<\sqrt{2\beta_0/\alpha}$.
So $\theta\leqslant 1-\sqrt{2\beta_0\alpha}-\theta_0 $ and thus $(1-\beta-\beta_0)-(1-\sqrt{2\beta_0\alpha}-\theta_0) =\sqrt{2\beta_0\alpha}+\theta_0-\beta-\beta_0<2\sqrt{d-1}/d$.
But this cannot hold together with (\ref{eq:necessary_large}) and (\ref{eq:necessary_small}).
So the condition required in (\ref{eq:necessary_propa}) is satisfied, and thus the conclusion holds.
\end{IEEEproof}

With this, the desired properties of the broadcast system can be directly supported upon $G$ with $\theta_0 n= (\beta_2-\beta_0)d+1$.
\begin{lemma}
\label{lemma_propagating2}
For any $\alpha,\beta,\beta_0,\beta_2\in (0,1)$, if
\begin{eqnarray}
\label{eq:necessary_large2}
\min\{\beta,\beta_2,1-\beta_2\}\geqslant\beta_0
\end{eqnarray}
and the inequalities (\ref{eq:necessary_large}) and (\ref{eq:necessary_small}) hold for $\theta_0= ((\beta_2-\beta_0)d+1)/n$, the $k_H$-\emph{Heaviside} and $k_\delta$-\emph{Dirac} properties are satisfied in $\mathcal{D}$ upon $G$ with bounded $k_H$ and $k_\delta$.
\end{lemma}
\begin{IEEEproof}
Firstly, if a node $j\in{P}(T,\beta_0)$ initiates a broadcast (as a \emph{General}) at $k_0$, as $j\in{P}(T,\beta_0)$ is a correct \emph{General}, there is $P_j\subseteq {P}(T,\beta_0)\cap N_j$ satisfying $|P_j|\geqslant(1-\beta_0)d+1$ and $\forall i\in P_j:x_i(k_0)=1$.
So with Lemma~\ref{lemma_propagating} we have $\forall i\in{P}(T,\beta_0), k\geqslant k_0+k_H: x_i(k)=1$ for some $k_H\leqslant |{P}(T,\beta_0)|$.

Next, if no node in ${P}(T,\beta_0)$ initiates any broadcast before $k_0$, as $\beta\geqslant\beta_0$ and there are less than $\beta_0 d$ $1$-neighbors of any node $i\in {P}(T,\beta_0)$ being out of ${P}(T,\beta_0)$, no such $i$ would be excited.
So as $\beta_2\geqslant\beta_0$, no node in ${P}(T,\beta_0)$ would be triggered before $k_0$.

Thirdly, if any node $i\in{P}(T,\beta_0)$ is triggered at $k_0$, there are at least $(\beta_2-\beta_0)d+1$ nodes in $N_i\cap{P}(T,\beta_0)$ are excited.
So with Lemma~\ref{lemma_propagating}, all nodes in ${P}(T,\beta_0)$ would be excited since $k_0+k_H$.
As every node $i\in {P}(T,\beta_0)$ has more than $(1-\beta_0)d$ $1$-neighbors in ${P}(T,\beta_0)$, with $\beta_2\leqslant 1-\beta_0$ every such $i$ would be triggered no later than $k_0+k_\delta$ for some $k_\delta\leqslant k_H+1$.
\end{IEEEproof}

As we can make $\mu< \sqrt{2\beta_0/\alpha}$, we would have $\lim_{n\to \infty}(\mu-1)\alpha/(1-\alpha)\leqslant\lim_{n\to \infty}(\sqrt{2\beta_0\alpha}-\alpha)/(1-\alpha)=0$ if $\alpha=n^{-\epsilon_1}$ for some $\epsilon_1>0$.
So by definition this $\mathcal{P}$ protocol upon $G$ is an a.e. $\epsilon$-incomplete protocol, providing that $\beta$, $\beta_0$ and  $\beta_2$ can be solved with (\ref{eq:necessary_large}), (\ref{eq:necessary_small}) and (\ref{eq:necessary_large2}).
Also, as the nodes need not know the network's actual topology, the propagation can run in dynamical networks, providing that the corresponding eigenvalues of the adjacency matrix of the continuously changing (and unknown) network are always sufficiently small.

Note, however, to satisfy (\ref{eq:necessary_large}), (\ref{eq:necessary_small}) and (\ref{eq:necessary_large2}), there are implicit limitations.
Firstly, as $1-\beta_2\geqslant\beta_0$ and $\beta\geqslant\beta_0$, we can set at most $\beta_2=1-\beta_0$ and at least $\beta=\beta_0$ in making rooms for setting $\beta_0$.
Secondly, by taking $\beta_2=1-\beta_0$, $\beta=\beta_0$ and $\theta_0 n= (\beta_2-\beta_0)d+1$ into (\ref{eq:necessary_small}) and then adding $((1-2\beta_0)d+1)/(3n)>\beta_0/3+\beta_0 -\sqrt{2\beta_0\alpha}$ to (\ref{eq:necessary_large}), we get $(d+1)/(3n)>\beta_0/3>2\alpha/3$ and thus $d+1>2\alpha n=2f$.
This means that the pure-propagation-based broadcast system can at best be built upon linear-degree networks.
In breaking this, the most trivial idea might be to enlarge the initial excitation area by directly adding extra edges to connect at least $s=\theta_0 n$ nodes for each node.
However, by doing this, the degrees of the nodes would also be increased to at least $\theta_0 n=O(t)$, which is still linear to $n$.
So we should make some further efforts to break this situation.

\subsection{Complementing a.e. propagation with localized communication}
\label{subsec:Complementing}
For sublinear-degree solutions, we look again to the a.e. propagation upon the Ramanujan network $G$.
The implicit linear-degree limitation mainly comes from $\theta_0$ being set as $((\beta_2-\beta_0)d+1)/n$, where the excitation of a very small area ($(1-2\beta_0)d$) is required to be propagated to almost everywhere of $G$.
From Lemma~\ref{lemma_propagating} we also see that if the initial excitation area can be somehow larger than $O(d)$, the condition on $\beta_0$ and $\beta$ could be much looser.
So the initial excitation area is the bottleneck of the fault-tolerant propagation.
Meanwhile, the advantage of fault-tolerant propagation is that, once the initial excitation area is sufficiently large, the cost of a.e. propagation is much lower than that of many other fault-tolerant communication protocols (such as the secure communication \citep{Upfal1992,Chandran2010}).
In a word, fault-tolerant propagation has the advantage of propagating to distant nodes when the propagated area is large.
While on the other side, many fault-tolerant communication protocols (including secure communication, Byzantine agreement, and so on) have the advantage of providing efficient fault-tolerance when the communication range is small.
So it is interesting to complement the advantages of distant-area propagation and nearby-region communication relatively.
Here, similar to the \emph{complementary filters} used in the frequency domain, we call such a relatively complemented $\mathcal{D}$ as a complementary system.

To construct a complementary system, we show that if $\theta_0$ can be sufficiently large, a.e. propagation can be reached in logarithmic time upon sublinear-degree networks.

\begin{lemma}
\label{lemma_sublinear_propa}
For every $d$-regular connected non-bipartite Ramanujan network $G$, if
\begin{eqnarray}
\label{eq:sufficient_large_theta}
\sqrt{d}> 4/(\theta_0+6\alpha-4\sqrt{2\alpha})
\end{eqnarray}
and
\begin{eqnarray}
\label{eq:sufficient_large_theta2}
\theta_0> \beta+\frac{3\beta_0}{1-\epsilon}-\frac{3-\epsilon}{1-\epsilon}\sqrt{2\alpha\beta_0}
\end{eqnarray}
hold for some constant $\epsilon\in(0,1)$, then there exists $\mathcal{D}$ upon $G$ such that for all $P_1\subseteq {P}(T,\beta_0)$ with $|P_1|\geqslant \theta_0 n$, if $\forall i\in P_1:x_i(k_0)=1$, then $\forall i\in{P}(T,\beta_0): x_i(k)=1$ holds for all $k\geqslant k_0+k_\delta$ with some $k_\delta=O(\log n)$.
\end{lemma}
\begin{IEEEproof}
With Lemma~\ref{lemma_propagating}, we need only to show $\beta_0$, $\beta$ can be solved with (\ref{eq:sufficient_large_theta}).
Concretely, to satisfy (\ref{eq:necessary_large}) and (\ref{eq:necessary_small}) with $\beta=\beta_0$, as $\beta_0+3(\beta_0-\sqrt{2\beta_0\alpha})<4\beta_0-6\alpha$, $\beta_0-\sqrt{2\alpha\beta_0}>\beta_0-\sqrt{2\alpha}$ and $\sqrt{d-1}/d<1/\sqrt{d}$, we need only to show $4(1/\sqrt{d}+\sqrt{2\alpha})\leqslant 4\beta_0<\theta_0+6\alpha$.
So $4(1/\sqrt{d}+\sqrt{2\alpha})<\theta_0+6\alpha$ would suffice.
For the worst-case propagation time, as the subgraph of $G$ induced by ${P}(T,\beta_0)$ is an expander, by extending the proof of Lemma~\ref{lemma_propagating} with (\ref{eq:sufficient_large_theta2}), with which we first suppose (and then get the similar contradiction) that there is only $S'\subset S$ satisfying $|S'|\leqslant (1-\epsilon)|S|$ and $\forall i\in S':|N_i\cap ({P}(T,\beta_0)\setminus S)|<\beta d$, we have $k_\delta=O(\log n)$.
\end{IEEEproof}

With Lemma~\ref{lemma_sublinear_propa}, if only $\theta_0>4\sqrt{2\alpha}-6\alpha$, there would exist a constant $d$ satisfying (\ref{eq:sufficient_large_theta}) for the $k_\delta$-round a.e. propagation.
Furthermore, it is easy to extend Lemma~\ref{lemma_sublinear_propa} to all $d$-regular strong enough expander $G$ with the second largest eigenvalue of the adjacency matrix of $G$ being $\lambda=O(d^{1/2})$ (see \cite{Upfal1992}, and other results for the explicitly constructed Ramanujan networks can also be extended similarly).
With this, the remaining problem is to construct some $s$-localized communication protocol $\mathcal{C}$ upon $G$ to support the desired $\theta_0$ with $s\ll n$.
Namely, with the $s$-localized $\mathcal{C}$ protocol running for some $s$-sized vertex-set of $G$, the end-to-end communication between the $s$ nodes in each such vertex-set would be localized.
Besides, it would be better if all the related communication paths in the localized communication protocol can also be localized in $O(\log s)$.
Further, it would be even better if all the end-to-end communication between the $s$ nodes can be accomplished between the same $s$ nodes.
Moreover, it would be optimal if all the related communication paths are with length $O(1)$.

In realizing the $s$-localized communication protocols, there can be different strategies.
Firstly, we can directly employ some incomplete secure communication protocol as $\mathcal{C}$.
With this, each node $i\in P(T,\beta_0)$ is expected to communicate with and only with up to $s$ nodes in $V$ (denoted as $S_i$, $i\in S_i$).
For efficiency, these $s$ nodes can be selected in the $c$-neighborhood of $i$ (with $c=O(\log n)$ in worst cases).
The $c$-neighborhood of $i$ is defined as $N_i^{(c)}=\cup_{r\leqslant c}L_i^{(r)}$, where $L_i^{(r)}=\{j\mid d_{i,j}^G=r\}$ is the set of all $r$-neighbors of $i$, with $d_{i,j}^G$ being the length of the shortest path between $i$ and $j$ in $G$.
Alternatively, we can also try to construct easier localized communication protocols other than secure communication.
As is limited here, we only discuss how to complement the a.e. propagation with the general $s$-localized communication protocol $\mathcal{C}$.

Firstly, for a.e. propagation, with Lemma~\ref{lemma_sublinear_propa}, it is desired that $\Omega(\alpha n)$ npc nodes should be initially excited.
Here we show that this can be satisfied by initiating the broadcast with the $s$-localized $\mathcal{C}$ protocol.
For this, we show that for every $T\in V_f$, there can always be $\Omega(\alpha n)$ npc nodes in some $c$-neighborhood of every npc node in $G$.

\begin{lemma}
\label{lemma_neighborhood}
If $G=(V,E)$ is a connected non-bipartite Ramanujan graph and (\ref{eq:necessary_large}) holds, then there exists $c=O(\log n)$ such that for every $T\in V_f$, $|N_i^{(c)}\cap P(T,\beta_0)|=\Omega(\alpha n)$ holds for all $i\in P(T,\beta_0)$.
\end{lemma}
\begin{IEEEproof}
As $G=(V,E)$ is a connected non-bipartite Ramanujan graph and (\ref{eq:necessary_large}) holds, for every $T\in V_f$, with Lemma~\ref{lemma_sufficient_large0} we have $|P(T,\beta_0)|\geqslant (1-\mu\alpha) n> (1-\sqrt{2\beta_0\alpha})n$.
With the proof of Lemma~2 of \cite{Upfal1992}, the subgraph of $G$ induced by $P(T,\beta_0)$, denoted as $G(T,\beta_0)$, is a (vertex) expander graph with an $a=\Omega((1/2-\beta_0)d)$ expansion coefficient.
Thus we have $d_{i,j}^{G(T,\beta_0)}=O(\log n)$ for all $i,j\in P(T,\beta_0)$.
As the constant $\beta_0<1/2$, we have $a=\Omega(d)$.
As $i\in P(T,\beta_0)$, we always have $|N_{i}^{(c+1)}\cap P(T,\beta_0)|\geqslant (a+1)|N_{i}^{(c)}\cap P(T,\beta_0)|$ for all $c\geqslant 0$ when $|N_{i}^{(c)}\cap P(T,\beta_0)|<n/2$.
So we have $|N_{i}^{(c)}\cap P(T,\beta_0)|\geqslant ((a+1)^{c+1}-1)/a=\Omega(d^{c})$.
So for every $i\in P(T,\beta_0)$ there is $c=O(\log n)$ such that $|N_i^{(c)}\cap P(T,\beta_0)|=\Omega(\alpha n)$.
\end{IEEEproof}

Now we show that there exists a.e. broadcast systems upon sublinear-degree networks by complementing the $s$-localized communication protocol and a.e. propagation.
\begin{theorem}
\label{theorem_broadcast_sublinear}
If there is an $s$-localized communication protocol $\mathcal{C}$ upon the $d$-regular $G$ with $s\geqslant u+\mu\alpha n$ and the premise of Lemma~\ref{lemma_neighborhood} holds, then a.e. broadcast system $\mathcal{D}$ exists upon some $d'$-regular $G'$ with $d'=d+O(1)$ with $k_\delta=O(\log n)$ and $k_H=O(\log n)$.
\end{theorem}
\begin{IEEEproof}
Firstly, with Lemma~\ref{lemma_neighborhood}, there exists a sufficiently large $c=O(\log n)$ such that for all $T\in V_f$, if $i\in P(T,\beta_0)$, then $|N_i^{(c)}\cap P(T,\beta_0)|=\Omega(\alpha n)$ holds.
So for every $T\in V_f$, if an npc-\emph{General} broadcasts at $k_0$, at least $\Omega(\alpha n)$ npc nodes would be excited before $k_0+O(\log n)$.
So with Lemma~\ref{lemma_sublinear_propa} we have $\forall i\in{P}(T,\beta_0): x_i(k)=1$ holds for all $k\geqslant k_0+O(\log n)$.
So by setting $D_y=H_{+u}$ with a sufficiently large $u=\mu\alpha n+4\sqrt{2\alpha}n$ and selecting $S_i\subset V$ with $s=u+\mu\alpha n$ for each $i\in V$, we have $\forall i\in{P}(T,\beta_0): y_i(k)=1$ holds for all $k\geqslant k_0+O(\log n)$.
And as no npc node would be excited if no npc-\emph{General} broadcasts in the underlying $\mathcal{P}$ protocol, the Heaviside property is satisfied.
For the Dirac property, if $y_i(k_0)=1$ holds for any $i\in{P}(T,\beta_0)$, we have at least $4\sqrt{2\alpha}n$ npc nodes being excited no later than $k_0$ in this case.
So again with Lemma~\ref{lemma_sublinear_propa} we have $\forall i\in{P}(T,\beta_0): x_i(k)=1$ for all $k\geqslant k_0+O(\log n)$.
And again with $s\geqslant u+\mu\alpha n$, $\forall i\in{P}(T,\beta_0): y_i(k)=1$ holds for all $k\geqslant k_0+O(\log n)$.
\end{IEEEproof}

\section{Conclusion}
\label{sec:con}
In this paper, we have investigated the broadcast problem upon bounded-degree networks with a simple but nontrivial system model.
In providing the relay-based broadcast systems upon bounded-degree networks, the a.e. propagation and the complementary systems are proposed upon strong enough expanders.
In building a.e. propagation upon the expanders, a general analysis of the fault-tolerant propagation is presented, and the related parameters are analysed.
In providing efficient broadcast systems, complementary systems are constructed by relatively complementing a.e. propagation and localized communication.
It is shown that by integrating a.e. propagation and localized communication protocols, more efficient broadcast systems can be built upon sublinear-degree networks than with only incomplete communication protocols.
This approach can go further to show to what extent the complexity of the Byzantine protocols and the node-degree of the networks can be lowered.
With the result of this paper, this mainly depends on the efficiency of the localized communication protocols.

\bibliographystyle{IEEEtran}
\bibliography{IEEEabrv,ASBBDNFTP}

\end{document}